\documentclass[11pt,preprint]{aastex}

\newcommand{\etal}{{\it et~al.}}
\newcommand{\ltsimeq}{\raisebox{-0.6ex}{$\,\stackrel{\raisebox{-.2ex}{$\textstyle <$}}{\sim}\,$}}

\begin{document}

\title{Revising the age for the Baptistina asteroid family using WISE/NEOWISE data}

\author{Joseph R. Masiero\altaffilmark{1}, A. K. Mainzer\altaffilmark{1}, T. Grav\altaffilmark{2}, J. M. Bauer\altaffilmark{1,3}, R. Jedicke \altaffilmark{4}}

\altaffiltext{1}{Jet Propulsion Laboratory/California Institute of Technology, 4800 Oak Grove Dr., MS 321-520, Pasadena, CA 91109, USA, {\it Joseph.Masiero@jpl.nasa.gov}}
\altaffiltext{2}{Planetary Science Institute, 1700 East Fort Lowell, Suite 106, Tucson, AZ 85719-2395}
\altaffiltext{3}{Infrared Processing and Analysis Center, California Institute of Technology, Pasadena, CA 91125 USA}
\altaffiltext{4}{Institute for Astronomy, University of Hawaii, Honolulu, HI 96822 USA}

\begin{abstract}

We have used numerical routines to model the evolution of a simulated
Baptistina family to constrain its age in light of new measurements of
the diameters and albedos of family members from the Wide-field
Infrared Survey Explorer.  We also investigate the effect of varying
the assumed physical and orbital parameters on the best-fitting age.
We find that the physically allowed range of assumed values for the
density and thermal conductivity induces a large uncertainty in the
rate of evolution.  When realistic uncertainties in the family
members' physical parameters are taken into account we find the
best-fitting age can fall anywhere in the range of $140-320~$Myr.
Without more information on the physical properties of the family
members it is difficult to place a more firm constraint on
Baptistina's age.

\end{abstract}

\section{Introduction}

The Main Belt asteroids (MBAs) offer a laboratory to study the
dynamical and collisional evolution of the inner Solar system, as well
as a window into the composition and thermal history of the protosolar
disk.  For nearly a century, asteroids grouped closely in orbital
element-space have been recognized as having formed from the
catastrophic disruption of a single larger parent body
\citep{hirayama1918,zappala90}.  Through modeling of the dynamical and
the non-gravitational forces that evolve the orbits of the family
members, the time since the breakup of the parent body has been
estimated.  The forces and processes that act on these small MBAs
depend on the bodies' physical parameters, such as diameter and
albedo.  Previous modeling methods have used the absolute visible
magnitudes of the family members as a proxy for their diameters
\citep[e.g.][]{nesvorny05}; however, this instills uncertainty in the
age determination as the derived age will depend strongly on the
assumed albedos.  Assumptions about other thermophysical parameters
will likewise introduce accompanying errors on the age determination.

The chronology of asteroid family breakups is one of the few methods,
along with cratering records and petrology/radioisotope ages, for
dating the history of events in the Solar system.  These collisional
events in the Main Belt can be linked to the geological record of the
Earth, as well as impacts on the terrestrial planets, other asteroids,
and the Earth's Moon
\citep[e.g.][]{delloro02,obrien05,farley06,cuk10,lefeuvre11}.  Ultimately, the
goal of such analyses is to understand the sequence of events in the
Main Belt and near-Earth object (NEO) populations that are known to
have had major consequences for life on Earth
\citep[e.g.][]{alvarez80}.  Finally, probing the ages of the oldest
families gives us a window into the most ancient history of the Solar
system, as some family formation events may coincide or even predate
the Late Heavy Bombardment and the epoch of giant planet migration in
the Solar system \citep{levison01,tsiganis05,morbi10dps,morbi10}.
 
Until recently, diameter measurements were only available for a few
thousand asteroids, most of these coming from the {\it Infrared
  Astronomical Satellite} ({\it IRAS}) survey \citep{tedesco02}.  With
the completion of the next-generation all-sky thermal infrared survey
by the Wide-field Infrared Survey Explorer \citep[WISE,][]{wright10}
and the identification of the small bodies of the Solar system
observed during that survey \citep[the NEOWISE project,][]{mainzer11}
a new data set has been opened.  NEOWISE allows us to determine
accurate diameters for the $>158,000$ observed Main Belt asteroids
detected during the fully cryogenic portion of the WISE mission and
albedos for the $>120,000$ that had previous optical measurements, of
which more than $33,000$ are members of previously identified asteroid
families \citep{masiero11}.  We can use these measured diameters of
family members to better constrain the ages of asteroid families by
revising predictions of their orbital evolution, using the methods
described in \citep{vok06}.

However, an important consideration in any attempt to determine
asteroid family age is the error introduced in that determination by
the assumed values of physical and orbital parameters.  Many physical
parameters (e.g. macroscopic density) are only poorly constrained for
more than a handful of objects, yet they play a large role in the
evolution of said bodies.  Similarly, the orbital parameters of the
parent body at the moment of breakup can only be assumed for families
older than a few million years \citep[cf.][]{nesvorny04}.

In this work we address both the uncertainty due to the assumed
initial conditions and the effect of using the newly available
diameter and albedo data from NEOWISE to the age determination of the
Baptistina asteroid family, using the work of \citet{bottke07} as a
starting point and road map.  In Section~\ref{sec.sim} we discuss the
numerical routines used to model the orbital evolution, as well as the
equations governing the thermal forces also acting on the body.  In
order to test the effect of the initial conditions chosen, we use the
assumed orbital and physical parameters from \citet{bottke07} and vary
each independently through a range of realistic values looking for
changes in the fitted age from their best-fit value.  We discuss the
behavior of the fit with respect to each of these parameters in
Section~\ref{sec.err}.  With these effects quantified, we can then
model the evolution of the family using the NEOWISE diameters and
albedos.  We discuss the new age determination in
Section~\ref{sec.newBap} and its implication in
Section~\ref{sec.conc}.

\section{Simulating Orbital Evolution}
\label{sec.sim}

Under the assumption of a common location and time of origin for the
members of a family, we can simulate the evolutionary history of the
orbits of family members using a numerical integrator.  For the Main
Belt, the dominant force shaping this evolution is the gravity from
the major bodies of the Solar system, in particular the Sun and
Jupiter.  However, non-gravitational effects such as those arising
from thermal radiation by the body can play an important role,
particularly for the smallest MBAs.  We discuss these two evolutionary
forces below in the context of the software used to model them.

\subsection{SWIFT}

The dynamic evolution of minor planets due to gravitational
interaction with the Sun is simulated using the Regularized Mixed
Variable Symplectic integrator as implemented in the SWIFT code
package \citep{levison94}.  This symplectic integrator calculates the
motion of a test particle by separating its Hamiltonian into two
parts: the Keplerian motion and the motion due to gravitational
interaction with other bodies, each of which can be solved
analytically.  One Hamiltonian is applied for half a time step, the
other is applied for the full time step, and the first is then applied
for the remaining half-step.  The Hamiltonian governing the
interaction acts as an acceleration in the particles' velocity, a
feature that is expanded on below when non-gravitational forces are
included.  This method of integration ensures that the energy of the
system is conserved.

SWIFT also includes the ability to handle close-approach cases between
particles at a much higher time resolution than is used for the
integration in general.  However, we have neglected this component of
the routine to reduce total run time.  As cases of
close-approaches/impacts with massive bodies will remove objects from
families instead of evolving them within the nominal orbital element
space, this assumption will not result in a significant increase in
the uncertainty of the family age.  We note that (as discussed below)
we do include the effect of non-destructive collisions on the
reorientation of the spin states and periods of the test bodies.

Required inputs for SWIFT are the initial positions of the test
particles (assumed to be all the same and coincident with the current
location of the parent fragment), the initial diameters ($D$), and the
initial velocities relative to the parent.  Each of the three velocity
components were assigned randomly up to a maximum value that is one of
the tested parameters ($V_0$) and scaled inversely proportionally to
the diameter of the body.  For this work we used a characteristic
diameter of $5~$km, following \citet{vok06}, to allow for comparison
with previous results.  We compare our simulations with the observed
family using two different methods of diameter determination
(depending on the goal of the simulation, as discussed below).  For
simulations that were compared to family lists generated from the
optically selected population (and thus without diameter information)
we used a single assumed albedo for the entire family and estimate
diameters from the $H$ absolute magnitude and the albedo.  For
comparisons to the families identified in the WISE data
\citep{masiero11} we use the diameters and albedos drawn from that
work.  Diameters from WISE were measured independently of other
sources of data, however the albedo measurements required a literature
$H$ magnitude and so are subject to optical observation biases and
errors.  It is important to note that the family lists used in
\citet{masiero11} were drawn from \citet{nesvorny06} who determined
family membership from a sample of optically-discovered asteroids; it
is expected that small, low albedo asteroids will be underrepresented
in these family lists, and that this may alter the determination of
family age.  Including asteroids discovered by WISE will begin to
mitigate this problem, and this will be the subject of future work.

\subsection{SWIFT\_RMVSY}
\label{sec.rmvsy}

To account for the non-gravitational forces due to thermal emission we
use the SWIFT\_RMVSY modification of the SWIFT code \citep{broz06}.
This upgrade uses the equations derived by \citet{vok98},
\citet{vok99a} and \citet{vok99b} to describe the thermal forces
acting on small Solar system objects.  When the thermal force modifies
the orbit of a body it is known as the Yarkovsky effect, and it occurs
when incident optical light is absorbed by a surface and re-emitted as
thermal infrared radiation in a different direction due to the
rotation of the body \citep[see][for a complete discussion]{bottke06}.
The Yarkovsky-O'Keefe-Radzievskii-Paddack (YORP) effect models the way
thermal radiation can change the spin state of non-spherical bodies
without atmospheres \citep{rubincam00}.

To calculate these thermal forces SWIFT\_RMVSY requires an input of
the thermal and physical parameters for each object: diameter, visible
geometric albedo ($p_V$), thermal conductivity ($K$), thermal capacity
($C_p$), infrared emissivity ($\epsilon$), surface density
($\rho_{s}$), bulk density ($\rho$), rotation rate ($\omega$), and
rotation pole orientation.  As a starting point for comparisons with
the WISE data, we assumed values of $K=0.01~$W m$^{-1}$ K$^{-1}$,
$C_p=680~$J kg$^{-1}$ K$^{-1}$, $\epsilon=1$, and
$\rho=\rho_s=2200~$kg m$^{-3}$, and assigned the population random
rotation rates and poles, following \citet{vok06}.  For comparisons
with literature work we use the same values assumed there.  We discuss
below the effects of varying these parameters on the best-fitting age.

As an object in the Main Belt evolves over time, it is predicted that
it will undergo small, non-disruptive impacts that can change the
body's rotation state (both spin pole and rotation period), occurring
with a characteristic timescale depending on diameter and rotational
angular momentum. We have modified the SWIFT\_RMVSY code to account
for this collisional reorientation by using the characteristic time of
reorientation ($\tau_r$) described by \citet{vok06}:
\begin{eqnarray}
\tau_{r} = B (\omega/\omega_0)^{\beta_1} (D/D_0)^{\beta_2}
\label{eq.reorient}
\end{eqnarray}
where $B=84.5~$kyr, $\beta_1=5/6$, $\beta_2=4/3$, $D_0=2~$m
\citep[i.e. a radius of $1~$m, see][]{farinella98}, and $\omega_0$
corresponds to a period of $5~$hr \citep[near the peak in the debiased
  distribution of MBA rotation rates, see][]{masiero09}.  In addition
to reorienting spin poles we also allow collisions to reset the
rotation rate of the body in a random fashion.

While collisional reorientation is treated as a random event, the
gradual reorientation of the spin axis by the YORP effect is treated
as a continuous change, preferentially driving the rotation pole
toward an asymptotic limit of $0^\circ$ or $180^\circ$ \citep{vok02}.
We use the median reorientation rate ($d\epsilon/dt=8.6~$deg/Myr) and
period doubling/halving time ($\tau_{per}=11.9~$Myr) derived from
thermophysical simulations of test bodies by \citet{capek04} for the
thermal conductivity matching our assumed value above ($K=0.01$).  We
note that we scale these timescales by the rotation rate as discussed
by those authors.  Following \citet{vok06} we also include a
multiplicative parameter $c_{YORP}$ that is applied to both YORP
parameters above ($\tau_{per}^\prime = \frac{11.9}{c_{YORP}}$ and
$d\epsilon/dt^\prime=c_{YORP}\times 8.6$) to model the uncertainty in
the age due to the weakly constrained YORP model.  This parameter has
been previously found to only show a weak effect on the age
determination \citep[cf.][]{vok06,bottke07} as long as it is
non-zero, though we discuss our findings further below.

\subsection{Supercomputing Resources}

Our numerical simulations make use of the supercomputing resources
available at NASA's Jet Propulsion Laboratory.  We used the Zodiac
supercomputer, comprised of 64 12-core Altix 2.66 Ghz nodes, for all
simulations discussed here.  Zodiac uses a 88 terabyte Lustre parallel
filesystem allowing for improved I/O capability, especially for rapid
writing to multiple files. Total peak performance is over 19
teraflops.  The range of simulations shown here required approximately
$300,000$ CPU hours of run time.

\subsection{Integration Step Size}

\begin{figure}[ht]
\begin{center}
\includegraphics[scale=0.5]{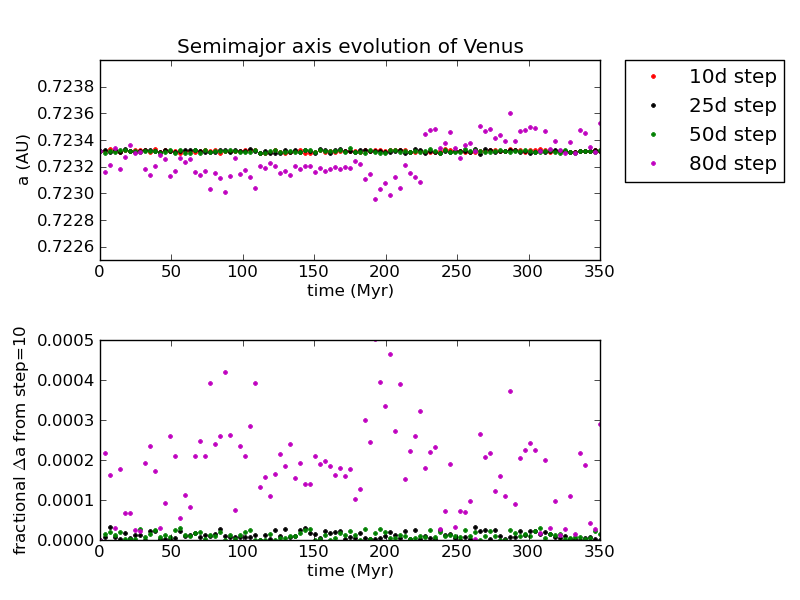}
\protect\caption{Simulated evolution of the orbit of Venus for varying
  integration step sizes.  For step sizes $\le50~$days there is no
  significant change in the semimajor axis that would indicate an
  increase in error due to an inappropriately large step size.}
\label{fig.step}
\end{center}
\end{figure}

For all the simulations we discuss in this manuscript, we included as
massive particles Venus, Earth, Mars, Jupiter, and Saturn, in addition
to the test particles and the Sun.  Uranus and Neptune are omitted as
they should play a much less significant role in the test particle
evolution than Jupiter and Saturn.  As Venus has the smallest
semimajor axis and perihelion of any tested body (with the exception
of MBAs ejected from the Belt into the NEO population, which are no
longer considered family members and hence are ignored once ejected)
our step size is restricted by Venus' orbital period.  It is
canonically recommended that the integration step size for a
symplectic integrator be $\ltsimeq 10\%$ of the period of the
innermost body (assuming a circular orbit) to prevent a rapid
accumulation of error on the total system energy
\citep[e.g.][]{broz06}.  We have tested the effect of step size on the
simulated evolution, and show in Figure \ref{fig.step} the semimajor
axis of Venus as a function of time for step sizes of 10, 25, 50, and
80 days, as well as the fractional change.  If the step size is
inappropriately large, we should see deviations in the the evolution
of Venus from the shortest time-step tested.  For step sizes
$\le50~$days we see no significant changes in the evolution of Venus
with respect to the 10-day step simulation.  For the remaining
simulations in this work we use a step size of $25~$days to ensure we
are well within the range of acceptable step sizes, finding it to be
the best balance between integration accuracy and time required to
perform the simulations.

\subsection{Family Membership}
\label{sec.members}

In order to determine the most accurate age possible for the family,
the list of family members that the simulations will be compared to
must have minimal corruption from asteroids that dynamically link to
the family but are not members.  This is a particular problem for the
Baptistina family, as the branch of the family that extends to smaller
semimajor axes overlaps with the much larger and older Flora family
\citep[cf.][]{nesvorny02}.

\begin{figure}[ht]
\begin{center}
\includegraphics[scale=0.5]{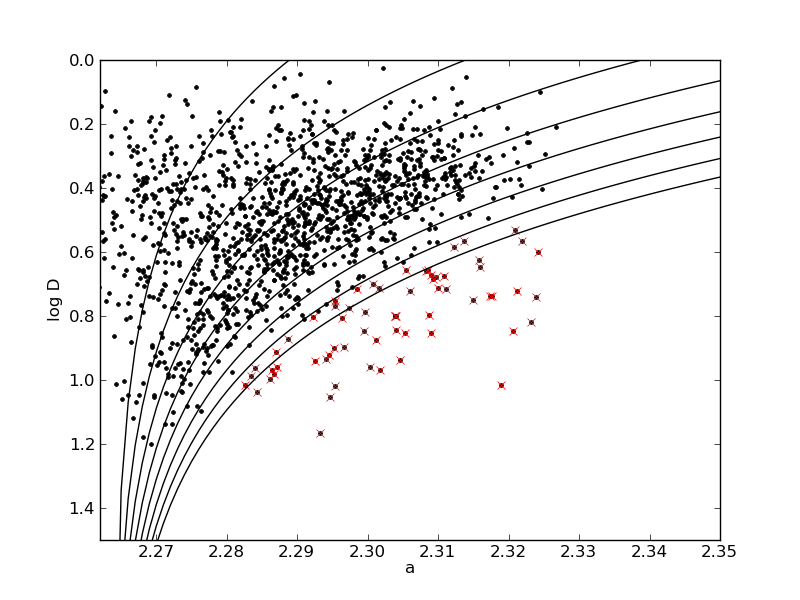} \protect\caption{Diameter
  vs. semimajor axis for the Baptistina family members used in our
  analysis.  The black lines show evenly-spaced steps of the
  C-parameter (see Section~\ref{sec.goodness}) used to compare family
  distributions, and points overlaid with a red `x' were assumed to be
  background objects and were not included in our analysis.}
\label{fig.cfig}
\end{center}
\end{figure}

Following \citet{bottke07} we restrict our analysis to consider only
the Baptistina family members at semimajor axes larger than the parent
body.  We have accomplished this by using the Hierarchical Clustering
Method \citep[HCM,][]{zappala90,zappala94} of family identification to
test a range of cutoff velocities.  We choose the highest velocity
that did not link to the lower-semimajor axis wing ($39~$m/s) as our
cutoff for family membership, following \citet{bottke07}.  Likewise,
we have removed from our list linked objects that are both large and
distant from the parent, and thus have a high probability of being
incorrect associations.  In Figure \ref{fig.cfig} we show the
resultant HCM-derived family that we use in our analysis.  Objects
that were rejected from the list are shown overlaid with an `x'.  We
note that while this will reduce uncertainty due to incorrectly
identified family members, it also decreases the sample size of
WISE-measured asteroids to $360$ objects and impacts our ability to
accurately compare the models to the true distribution.
Identification of new family members and measurement of their physical
parameters will help us decrease these uncertainties.

\subsection{Goodness of Fit Determination}
\label{sec.goodness}

We cannot uniquely link individual test particles to observed family
members as the randomized initial conditions will not necessarily mean
the evolutions are identical.  Instead we focus on the distribution of
the true and test populations to find the best matching initial
conditions.  To compare our simulation to the known population, we
perform a $\chi^2$ test of the $C$ parameter, which is defined as
$C=\Delta a~10^{-0.2 H}$ by \citet{vok06} for cases where the albedo
is unknown.  For tests conducted using only objects with physical
parameters measured by WISE, we define a $C_D$ parameter as
$C_D=\Delta a~D$ (where $D$ is the diameter) that is roughly
equivalent to the $C$ parameter with a multiplicative offset.  Larger
objects are predicted to have smaller drift rates from
non-gravitational effects, and so the $C$ and $C_D$ parameters
represent lines of constant time for a given drift strength.  An
important difference is that $C_D$ has no dependence on the albedo of
the asteroid, unlike $C$.

Figure \ref{fig.cfig} shows a series of curves indicating $C_D$ values
from $0.025$ to $0.2$ in steps of $0.025$ overlaid on the Baptistina
family.  To compare simulations to reality, we compare the $C$ or
$C_D$ distribution of the simulation to the same distribution for the
family.  As discussed above we only use family members at semimajor
axes larger than the parent, and thus likewise only consider simulated
particles that are in that same region of semimajor axis-space at the
timestep being tested.  We note that it is possible for a particle to
begin the simulation drifting inward and later through reorientation
begin moving outward.  Thus it is possible for that particle to be
used for the comparison to the observed family members at some
timesteps but not others.  The goodness of fit at each time step is
obtained from a bin-by-bin $\chi^2$ comparison of the two populations.
The match to the observed data initially improves as the test
bodies disperse over time, until they expand beyond the observed
population and the $\chi^2$ climbs.  The time at which the minimum
$\chi^2$ is reached is therefore the best-fit to the present day
family, and thus can be inferred to be the age of the family.

\section{Errors Due to Assumed Physical and Orbital Parameters}
\label{sec.err}

The numerical simulations of the orbital evolution of family members
are deterministic in the sense that the equations of motion (both
gravitational and non-gravitational) can be described analytically.
However, the specific behavior of an individual particle depends
strongly on the initial conditions assumed for it, including the
physical, orbital, and spin state parameters.  While the effect of the
randomized initial conditions on the behavior of the population should
fade as the population of test particles grows (e.g. the initial spin
pole and rotation rate, the randomized collisional reorientation of
particles, etc.), other initial conditions that are singularly chosen
for the population and do not change with time may have a dramatic
effect on the overall evolution.

Before attempting to fit ages for asteroid families, we first will
test our dependence on the chosen value for each parameter.  We
constrain the possible errors induced by assumptions of the thermal
parameters ($K$, $\epsilon$, $C_p$), orbital parameters (mean anomaly,
longitude of perihelion, longitude of the ascending node), and
physical parameters ($\rho$, spin state).  We include $V_0$ and
$c_{YORP}$ as tested parameters that are varied to find the
best-fitting age, and so will not discuss them here.  Additionally, it
is beyond the scope of the work presented here to investigate the
effect of varying the equations governing the velocity distribution of
the impact ejecta (here assumed to be $V=V_0 \frac{5 km}{D}$) and
collisional reorientation (Equation~\ref{eq.reorient}), however these
also will act as a source of uncertainty.

For the tests of the physical and orbital parameters, we follow the
assumed initial conditions and albedo for the Baptistina family from
\citet{bottke07} for the purpose of comparison.  Once the uncertainty
due to the assumed initial conditions has been quantified, we conduct
a new set of simulations that use the measured values for the
diameters and albedos in Section~\ref{sec.newBap} to update the age of
the Baptistina family.

Following the best-fit values from \citet{bottke07} for Baptistina, we
assume a breakup velocity for the parameter tests of $V_0 = 40~$m
s$^{-1}$, $c_{YORP}=1.0$, $K=0.01~$W m$^{-1}$ K$^{-1}$, $C_p=680~$J
kg$^{-1}$ K$^{-1}$, $\epsilon=1$, $\rho=\rho_s=1300~$kg m$^{-3}$, and
randomized rotation states.  We note that using identical initial
conditions we reproduce the best-fitting age of $T\sim160~$Myr for the
family found by those authors.  In order to compare our results
directly to previous work, we initially use the $H$ magnitudes along
with the assumed albedo used by those authors ($p_V=0.05$).  In
Section \ref{sec.newBap} we use the WISE measured diameters and
albedos.

\subsection{Rotation State}

The assumed initial rotation pole and period of a test particle will
dictate the magnitude and direction of the Yarkovsky force at the
outset of the simulation.  Over the course of the evolution of the
family, the YORP effect will gradually reorient the spin axis of a
test particle and slow or speed its rotation \citep{capek04}, while
collisions will occasionally abruptly randomize these values.  While
YORP, by driving the rotation poles to obliquities of $0^\circ$ or
$180^\circ$, will in general increase the magnitude of the Yarkovsky
effect, collisions are more likely to decrease its strength or reverse
it completely.

\begin{figure}[ht]
\begin{center}
\includegraphics[scale=0.5]{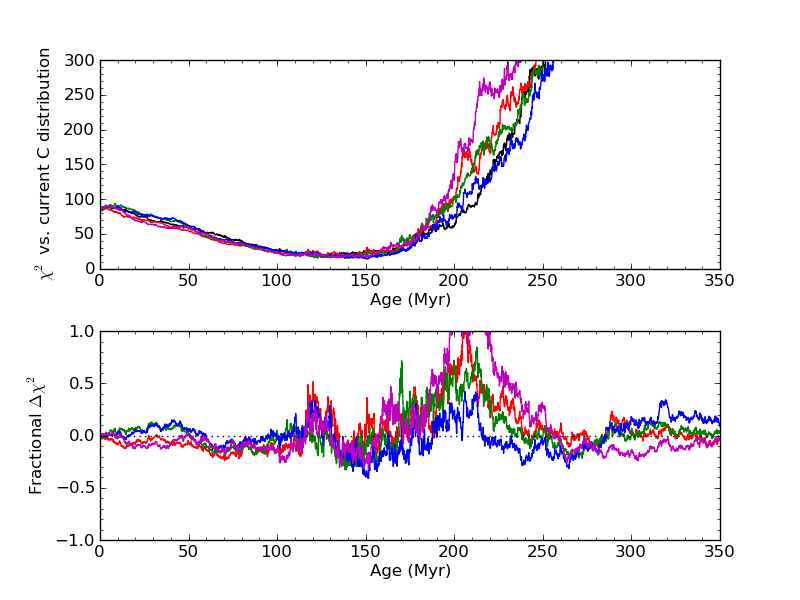}
\protect\caption{Identical simulations of the evolution of the
  Baptistina family changing the initial, random spin states of the
  test particles.  The lower plot shows the fractional difference
  between the first test and the other four, for comparison.}
\label{fig.spin}
\end{center}
\end{figure}

In Figure~\ref{fig.spin} we show five identical simulations of the
Baptistina family, allowing only the randomized spin states of the
test particles to vary.  The evolution of these simulations varies in
$\chi^2$ by $\sim25\%$ for the first $175~$Myr.  After this point,
when the comparisons between the simulations and the real distribution
become rapidly worse, the differences between simulations increases
however this regime is less deterministic of age of the family.  This
results in an uncertainty in the specific best-fit age of $\sim20~$Myr
in the case of Baptistina, however the range of likely ages remains
comparable.

\subsection{Thermal Properties}

The thermal parameters of the test particles can have a significant
effect on the evolution of the family.  We therefore have tested the
effect of altering the assumed thermophysical parameters on the
evolution of the test population.  In particular, we focus on varying
$\epsilon$, $K$, and $C_p$ across ranges typical for real-world
materials around the default assumed values of $\epsilon_0=1.0$,
$K_0=0.01~$W m$^{-1}$ K$^{-1}$, and $C_{p,0}=680~$J kg$^{-1}$
K$^{-1}$.

\begin{figure}[ht]
\begin{center}
\includegraphics[scale=0.5]{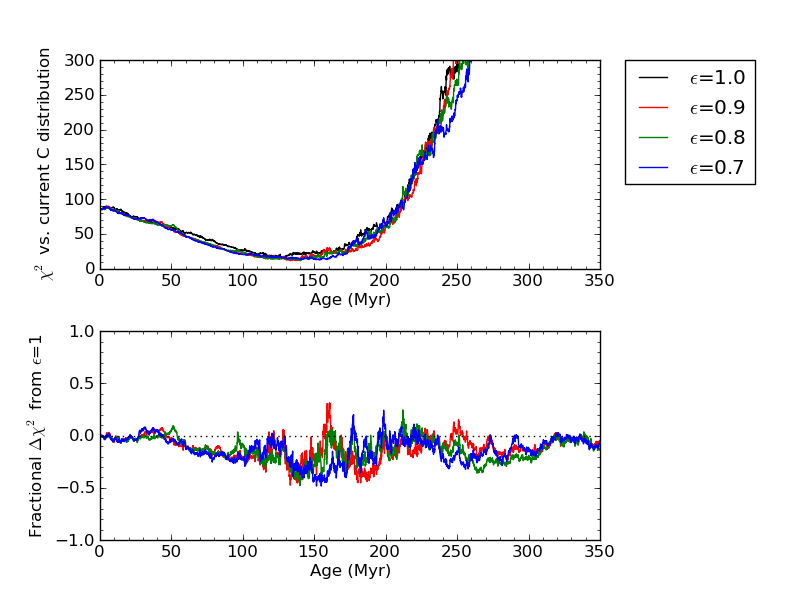}
\protect\caption{The same as Figure~\ref{fig.spin}, but now testing
  various values of emissivity ($\epsilon$).  The lower plot shows the
  fractional difference between the $\epsilon=1$ case and the other
  tests, for comparison.}
\label{fig.emis}
\end{center}
\end{figure}

\begin{figure}[ht]
\begin{center}
\includegraphics[scale=0.5]{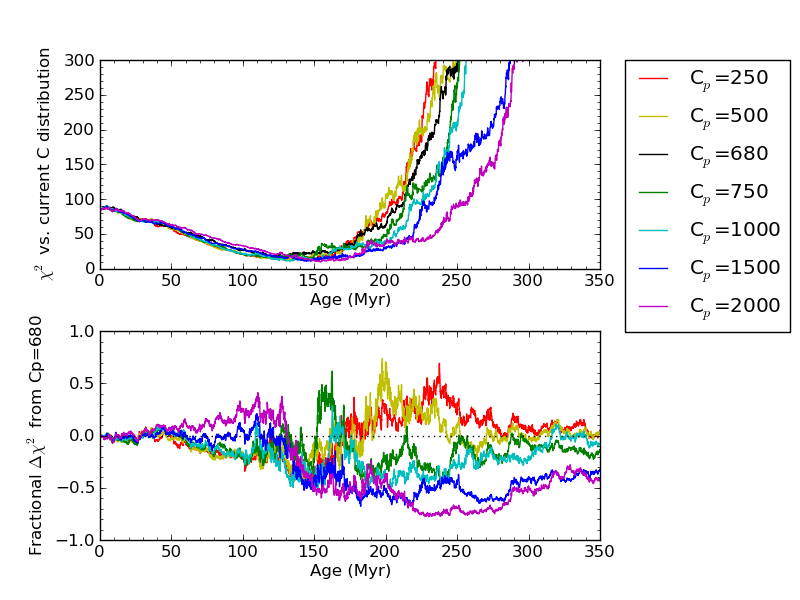} \protect\caption{The same
  as Figure~\ref{fig.spin}, but now testing various values of thermal
  capacity ($C_p$).  The lower plot shows the fractional difference
  between the $C_p=680~$J kg$^{-1}$ K$^{-1}$ case and the other tests,
  for comparison.}
\label{fig.Cp}
\end{center}
\end{figure}

We show in Figure~\ref{fig.emis} the evolution of the Baptistina test
family for various initial emissivity values, over the range of $0.7
\le \epsilon \le 1.0$.  We see no significant differences between each
of the cases when only thermal emissivity is varied.  Thus, our
assumed value for emissivity of $\epsilon=1.0$ is valid for future
tests.  Likewise, in Figure~\ref{fig.Cp} we show the evolution of the
test family comparing a wide range of different thermal capacities:
$250<C_p<2000~$J kg$^{-1}$ K$^{-1}$.  We again see no significant
changes at ages less than $150~$Myr.  Beyond this age, the simulations
appear to sort roughly corresponding to $C_p$, where simulations with
smaller values of $C_p$ diverge from the real population faster than
those with larger $C_p$.  For the purposes of finding the best fit
age, an assumed value of $C_p=680~$J kg$^{-1}$ K$^{-1}$ is adequate.

\begin{figure}[ht]
\begin{center}
\includegraphics[scale=0.5]{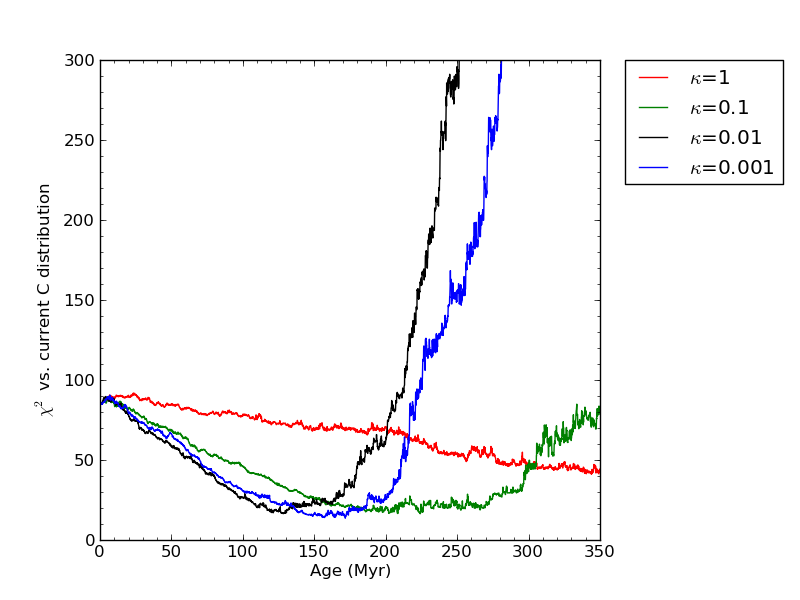}
\protect\caption{The same as Figure~\ref{fig.spin}, but now testing
  various values of thermal conductivity ($K$).  }
\label{fig.cond}
\end{center}
\end{figure}

Conversely, we find that the assumed value of thermal conductivity
($K$) has a significant impact on the strength of the thermal forces
acting on the bodies, as it is the only parameter that varies over
many orders of magnitude in realistic materials.  \citet{vok98} show
in their Figure 3 the relative strength of the transverse Yarkovsky
force vector as a function of the thermal parameter $\Theta$; using
$K=0.01~$W m$^{-1}$ K$^{-1}$ and the nominal assumptions for $C_p$,
$\epsilon$, $\rho$, and rotation rate places $\Theta$ at the peak
value for the transverse force.  Changes in $K$ by half or one order
of magnitude result in a significant change in the strength of the
Yarkovsky effect.  Following the thermal inertias ($\Gamma$) found by
\citep{delbo09} for asteroids with $D<200~$km, we test a range of
thermal inertia values of $40<\Gamma<1200$J s$^{-0.5}$ m$^{-2}$
K$^{-1}$ which corresponds to thermal conductivities of $0.001<K<1$
for nominal values of density and thermal capacity.  We show the
results of these simulations in Figure~\ref{fig.cond}.  The evolution
of the test family is significantly slower for values both larger and
smaller than $K=0.01~$W m$^{-1}$ K$^{-1}$.  We note that while
$K\sim1$ is only observed for the smallest of near-Earth asteroids
that are believed to have surfaces free of regolith and thus may not
be a good analog for $D\sim5~$km MBAs, the range of $0.001<K<0.1$ is
still possible for MBAs.  We use $K=0.01~$W m$^{-1}$ K$^{-1}$ for
future simulations, however this probably represents only a lower
limit on the family age.  Determination of thermal conductivity or
thermal inertia for a number of family members will be critical to
determining the true evolution of the family.

\subsection{Initial Orbit}
\label{sec.orbit}

In order to model the breakup of a family, we assume that all members
began at the same place in space and time, and assign them an ejection
velocity that scales inversely with their diameter
\citep[following][]{vok06}, which combines with the particle's
velocity around the sun to generate a new orbit.  As the ejection
velocities typically are small compared to the motion around the sun,
this will preferentially elongate the cloud along the path of the
orbit.  Although the velocity imparted on the fragments by the
collision will be randomized around a constant value, for a parent
body with an eccentric orbit the change in orbital parameters after
the impact can vary depending on the parent's mean anomaly at
the time of breakup.  Nominally we use the present day osculating
orbital elements for the largest family member as the orbit of the
body prior to breakup, ensuring that the test particles are in the
same osculating system as the planets (including using the same
assumed epoch).  However, we have tested the results of varying the
mean anomaly (MA), longitude of perihelion ($\varpi$), and longitude of the
ascending node ($\Omega$) on the subsequent evolution of the family.

\begin{figure}[ht]
\begin{center}
\includegraphics[scale=0.5]{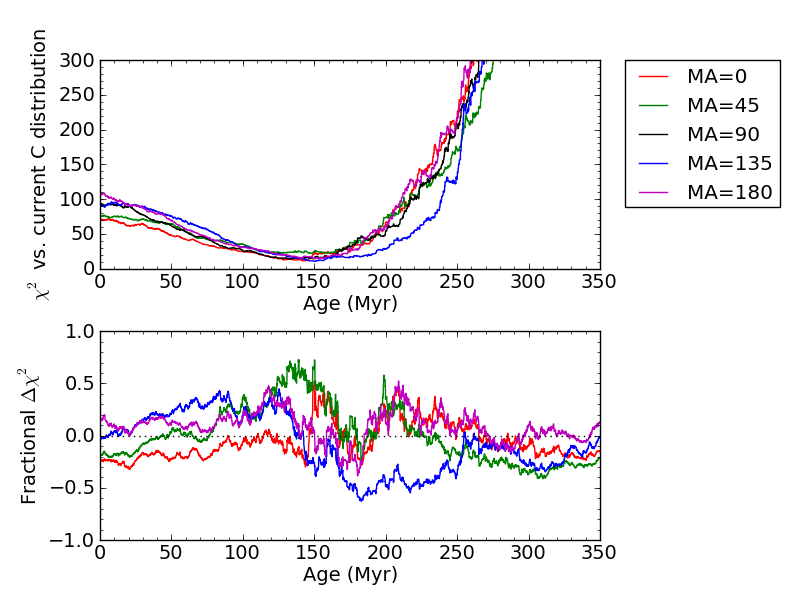} \protect\caption{The same as
  Figure~\ref{fig.spin}, but now testing a range of Mean Anomaly
  values at the time of breakup.  }
\label{fig.MA}
\end{center}
\end{figure}

In Fig~\ref{fig.MA} we show a range of simulations with identical
physical parameters, $c_{YORP}$ and breakup velocity $V_0$, while
stepping through mean anomaly of the parent at the time of breakup.
The velocity added to a test particle's motion upon breakup alters its
initial orbit.  However, the initial impulse is more effective at
changing the orbit's aphelion when the breakup is at perihelion than
it is at changing the orbit's perihelion when the breakup is at
aphelion.  This effect is shown in Figure~\ref{fig.MA} as the offset
in $\chi^2$ at $T=0$, where simulations with breakups closer to
perihelion have a larger initial spread in semimajor axis and thus a
lower $\chi^2$.

In general, after about $\sim 100~$Myr the differences between
populations with different initial mean anomalies are erased by the
effect of Yarkovsky-induced drifts and gravitational orbital
evolution.  We note that this timescale will depend on the initial
eccentricity of the parent body: parents with low or zero eccentricity
should see little difference in family member distribution between
breakups at perihelion or aphelion even at $T=0$, while those with
larger eccentricities will require more time to erase the initial
differences.  This effect may thus be particularly important for
high-eccentricity families younger than $\sim100~$Myr.

\begin{figure}[ht]
\begin{center}
\includegraphics[scale=0.5]{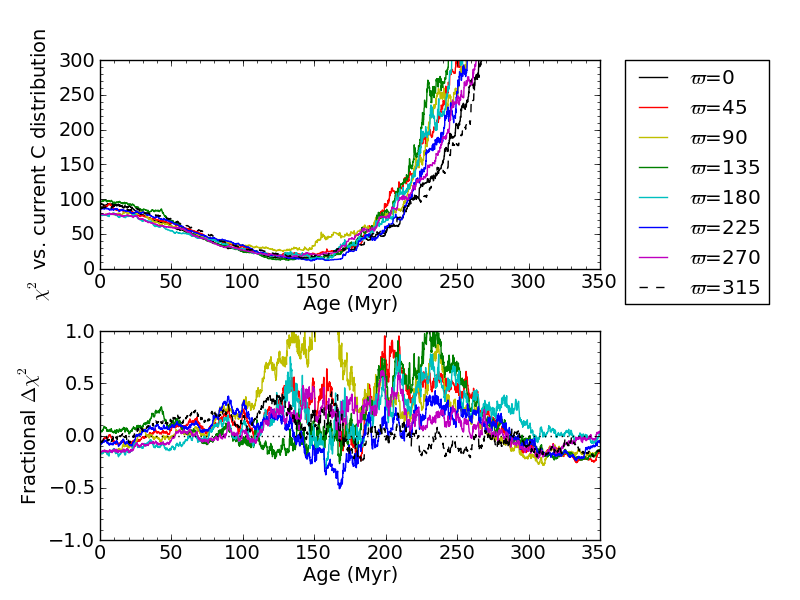} \protect\caption{The same
  as Figure~\ref{fig.spin}, but now testing a range of values for the
  longitude of perihelion ($\varpi$) at the time of breakup.  }
\label{fig.peri}
\end{center}
\end{figure}

\begin{figure}[ht]
\begin{center}
\includegraphics[scale=0.5]{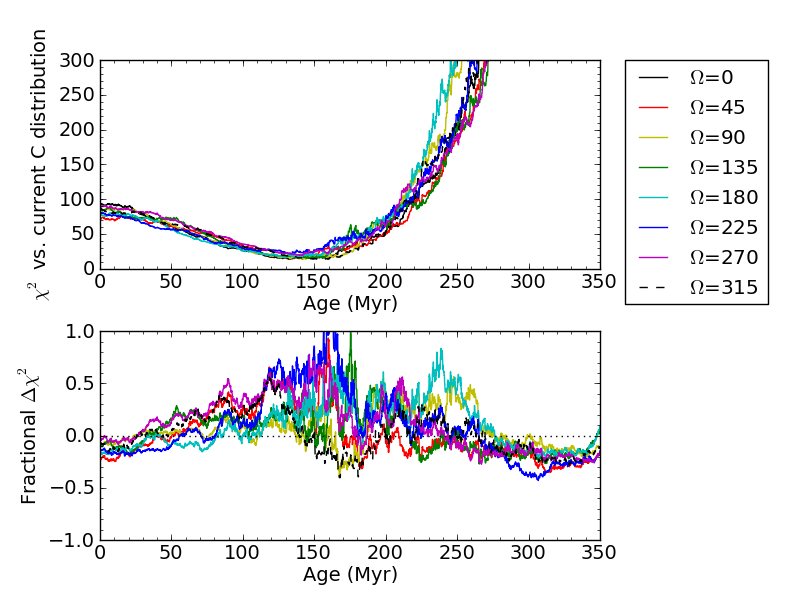} \protect\caption{The same
  as Figure~\ref{fig.spin}, but now testing a range of values for the
  longitude of the ascending node ($\Omega$) at the time of breakup.
}
\label{fig.node}
\end{center}
\end{figure}

We show in Figures \ref{fig.peri} and \ref{fig.node} the results of
similar simulations, testing $\varpi$ and $\Omega$ respectively.
Variations in both parameters result in no significant change to the
evolution of the population in general.  While these parameters may
have an effect for other families with parents significantly more
eccentric than Baptistina, we can safely use the present-day
osculating values for all parameters for the simulations we discuss in
Section \ref{sec.newBap}.

\subsection{Density}
\label{sec.density}

A key assumption in determining the strength of the Yarkovsky effect
on the orbit of an asteroid is the mass of the body.  The Yarkovsky
effect is expected to produce a force that depends on the illuminated
area of the body, but the resultant acceleration will scale with the
mass.  While the SWIFT\_RMVSY code includes a parameter to allow for
testing the variation in the strength of the YORP effect due to the
uncertainty in its absolute strength (the $c_{YORP}$ parameter), the
Yarkovsky effect should be well quantified if the mass and thermal
parameters are known and thus does not include this scaling parameter.
With WISE we can usually derive effective diameters to within
$\sim10\%$ for asteroids observed with good signal-to-noise
\citep{mainzer11cal}.  However the bulk density of asteroids remains
poorly constrained.  Likewise surface density, which is a component of
the calculation of thermal propagation in Yarkovsky, is equally
difficult to determine and is assumed here to be equal to the bulk
density.

Density measurements of meteorites can provide an upper limit to the
density we expect for different compositions of asteroid, but linking
meteorites to asteroids can be difficult, and the macro- and
micro-porosity of a body (which will strongly affect the measured bulk
density) are almost impossible to measure remotely.  Conversely,
asteroid masses can be obtained from their gravitational perturbation
of the other objects in the Main Belt (for the few largest bodies),
from deviations on spacecraft trajectories during fly-by (for the
handful of objects visited by spacecraft), or from the periods of
satellite bodies in orbit around the asteroid of interest (if
satellites are known to exist and the periods can be measured).

\citet{carry12} provides a thorough review of the state of knowledge of
asteroid densities.  They list densities for $38$ MBAs smaller than
$D=200~$km and with density accuracy better than $20\%$.  The mean
density of this group is $\rho=2.3\pm1.2~$g cm$^{-3}$, however the
error is inflated by the range of compositions.  Attempting to trace
composition with spectral taxonomy, they show that the range of bulk
density within a given spectral taxonomic class can still be large,
due to changes in macroporosity which they attribute to increasing
compaction at larger diameters.  The authors show some correlation
between density and spectral type (though even then the intrinsic
scatter is about $\sim25\%$ in the best cases) and find similar
discrepancies to the ones seen by \citet{mainzer11tax} when comparing
taxonomy and albedo, notably for the objects spectrally identified
with the X-complex.  Without an independent measurement of the density
of a significant number of family members, age fits must be performed
over the entire viable range of bulk densities.  Otherwise, improperly
narrow error windows on the best-fit age will be derived.  If the
family taxonomy can be linked to meteorite analogs, a smaller window
can be used, though the unknown porosity will still induce uncertainty
in the density estimate.

\begin{figure}[ht]
\begin{center}
\includegraphics[scale=0.5]{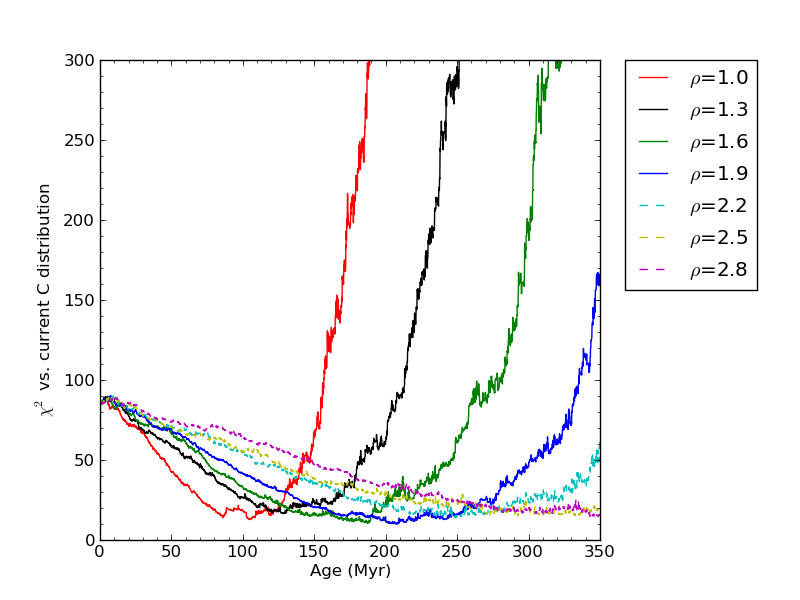}
\protect\caption{Simulations of the evolution of the Baptistina family under varying assumptions for the bulk and surface density of the test particles (assuming both densities are equal).}
\label{fig.density}
\end{center}
\end{figure}

We show in Figure~\ref{fig.density} simulations of evolution of the
Baptistina family, in this case only varying the density assumed for
the family members over a range of $1.0< \rho <2.8~$g cm$^{-3}$.  All
other physical and orbital parameters follow the assumptions used in
Section~\ref{sec.orbit}.  The rapid change in best-fitting age for
different densities is a result of the weakening of the accelerative
kicks in the orbital velocity from the Yarkovsky force (i.e. for an
assumed diameter the force will be constant, while the acceleration
will be inversely proportional to the mass and thus the density).
This is shown by the $\chi^2$ minimum best-fitting age $T$ following a
general $T \propto \rho^{-1}$ where $\rho \ltsimeq 2$, above which the
best-fitting age increases rapidly.

\citet{bottke07} assumed that both the bulk density and surface
density of the Baptistina family members were $1.3~$g cm$^{-3}$ from
their assumption that the spectral taxonomy was most similar to a
C-type asteroid \citep[however see][for further discussion on the
  taxonomy of Baptistina and its family]{reddy09,reddy11}.  For our
revised simulations (see Section~\ref{sec.newage}) we adopt a bulk and
surface density of $2.2~$g cm$^{-3}$, assuming S-type taxonomy.
However testing over the full range of probable densities ($\sim1.6$
to $\sim2.8$) will result in a broadening of the best fit range.

\section{The Age of Baptistina Incorporating WISE Results}
\label{sec.newBap}

Using the methodology developed by \citet{vok98} we revise the
estimated age of the Baptistina family by \citet{bottke07} by taking
into account the diameter and albedo measurements offered by NEOWISE
for $\sim1/3$ of the known family members.  One complicating factor in
identifying and modeling this family is its partial overlap in orbital
element space with the much larger and older Flora family.  The albedo
distinction between these two families should enable us to use this
parameter as a further restriction on family membership, and
development, testing, and analysis of this method will be presented in
a future paper.  For this preliminary analysis we use a restricted set
of family members that includes only the objects that have drifted
outwards from the parent and thus are not contaminated by Flora, as
discussed above in Section \ref{sec.members}.

\subsection{New Observational Data}

The diameters and albedos we use for this work are drawn from the
values derived for MBAs published in \citet{masiero11}.  The larger
asteroids were more likely to have been seen in multiple bands by WISE
which allows for fitting of the beaming parameter.
\citet{mainzer11cal} show that in cases such as this the absolute
error on diameter is $\sim10\%$ and on albedo is $\sim20\%$ of the
measured albedo value, however internal comparisons are better than
this limit.  We note that this albedo error assumes moderate-to-low
light curve amplitudes and well characterized H and G values.  In
addition to observing known objects, NEOWISE also discovered new
asteroids, preferentially with lower albedos where ground-based
surveys are less sensitive.  These previously unknown objects
represent a source of error in the diameter and albedo distribution of
known families as they would not be included in the known family
lists, and will tend to make the true albedo distribution darker than
the distribution seen for the previously known asteroids, most of
which were discovered by visible light surveys that are biased against
detecting low albedo objects.  Future work will address the error
resulting from this change in albedo distribution.

The primary variation between the observed data and the assumed values
in \citet{bottke07} is the average value for the albedo of the family
measured by WISE ($p_V=0.21$) compared with the assumed value used
previously ($p_{V,assumed}=0.05$).  The main effect of this change is
to reduce by more than a factor of two the effective size of a typical
Baptistina family member used in simulations.  We note that because
the albedo distribution of the Baptistina family is fairly wide
($\pm0.1$), the change in diameter from assumed to measured values for
each individual family member can be much larger or smaller than the
factor of two derived from applying the mean albedos.  It is therefore
critical to use the actual measured diameters for family members where
available, instead of assuming a uniform albedo for all objects.  This
will also remove an additional source of uncertainty that is inherent
to the $H$ magnitude measurement.

As discussed in Section \ref{sec.density}, the density chosen for the
test particles can have a very large effect on the best-fit age that
is determined for the family.  \citet{bottke07} use a density of
$1.3~$g cm$^{-3}$ appropriate for small C-complex bodies
\citep{carry12}, as Baptistina was thought to be.  The revisions in
asteroid sizes and albedos from the WISE data, as well as taxonomic
classification of a larger set of Baptistina family members as
S-complex bodies \citep{reddy11} drives us to assume a larger bulk
density for the objects.  For our initial simulations, we assume
$\rho=2.2~$g cm$^{-3}$, however we also test a range of densities
using the updated diameters.

\subsection{Revised Age and Error}
\label{sec.newage}

\begin{figure}[ht]
\begin{center}
\includegraphics[scale=0.6]{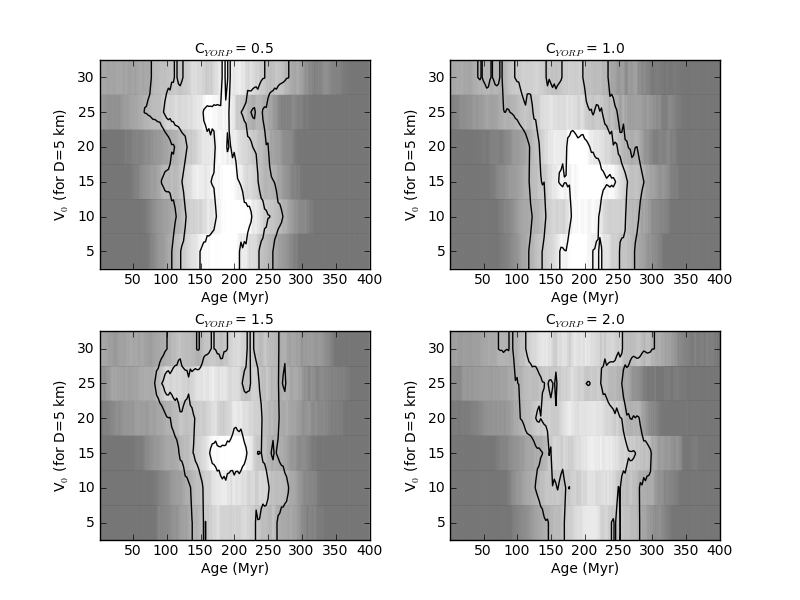}
\protect\caption{$\chi^2$ maps of breakup velocity $V_0$ vs. age for
  the four tested values of $c_{YORP}$, with white shading
  representing the best fits and dark shading the worst.  Contours
  show $\chi^2$ levels of $6, 12, 18$, the first of which defines the
  boundary of the region of acceptable fits.}
\label{fig.newageC}
\end{center}
\end{figure}

\begin{figure}[ht]
\begin{center}
\includegraphics[scale=0.7]{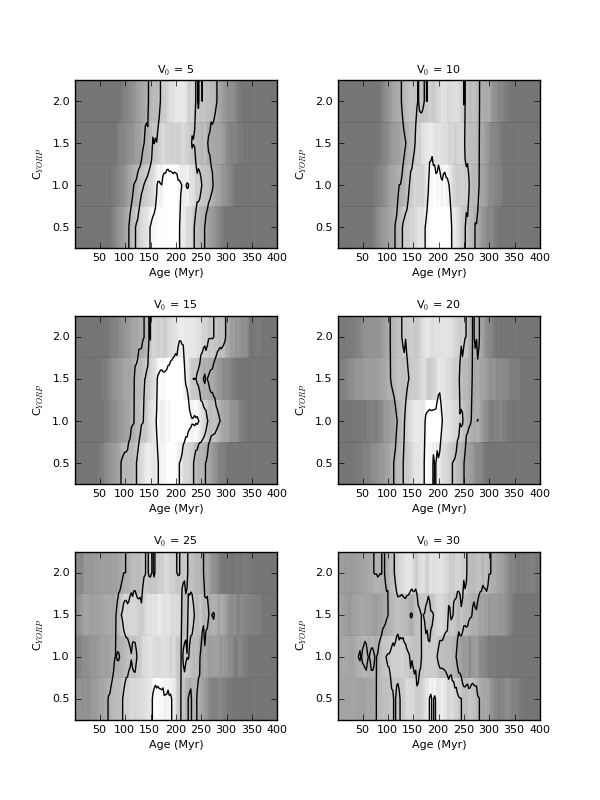} \protect\caption{The
  same as Figure \ref{fig.newageC} but showing $c_{YORP}$ vs. age for
  the six tested values of $V_0$.}
\label{fig.newageV}
\end{center}
\end{figure}

Using a set of test particles with the same size and albedo as were
measured for the Baptistina family by WISE, we simulate their
evolution over $400~$Myr using for our initial conditions: present day
osculating elements for Baptistina and Venus through Saturn,
$\epsilon_0=1.0$, $K_0=0.01~$W m$^{-1}$ K$^{-1}$, $C_{p,0}=680~$J
kg$^{-1}$ K$^{-1}$, and $\rho_0=\rho_{s,0}=2200~$g cm$^{-3}$.  We
initially test a grid of breakup velocities ($V_0$) and $c_{YORP}$
parameters.  In Figure \ref{fig.newageC} we show $\chi^2$ maps of
$V_0$ vs. age for each of the four tested $c_{YORP}$ values.  Figure
\ref{fig.newageV} shows an alternate view of the same simulations,
with each map showing $c_{YORP}$ vs. age for a given $V_0$.  For this
limited range of assumed parameters, the best fit age is
$190\pm30~$Myr, with minimal dependence on $V_0$ and $c_{YORP}$ in the
ranges of $5<V_0<20~$m/s and $0.5<c_{YORP}<1.5$, and only a slight
preference for lower values in each case.  We note that due to the
albedo assumed by \citet{bottke07} of $p_V=0.05$, the inferred
diameters are approximately a factor of two larger for the family
members and thus it is not unexpected that their best-fit $V_0\sim40$
is similarly larger than the best-fit value we find.

\clearpage

\begin{figure}[ht]
\begin{center}
\includegraphics[scale=0.6]{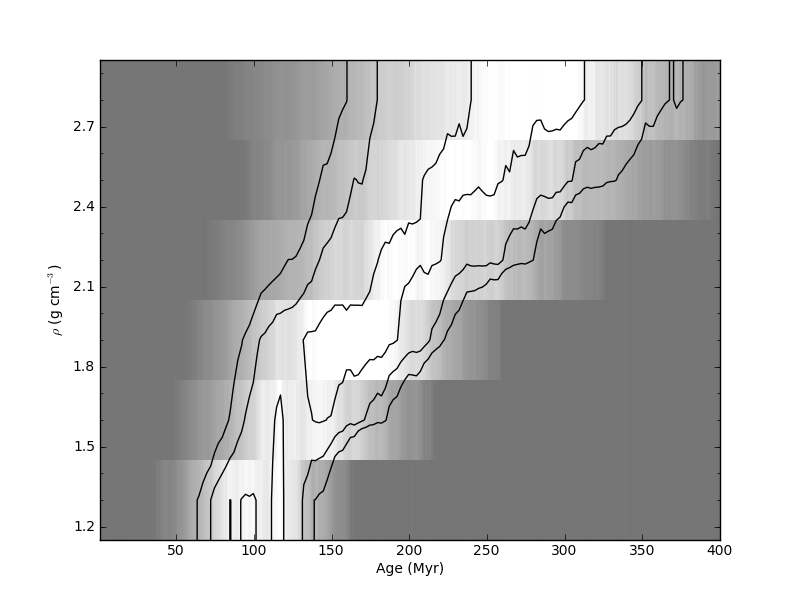}
\protect\caption{The same as Figure \ref{fig.newageC} but showing
  density $\rho$ vs. age assuming the best fit values of $V_0=10~$m/s
  and $c_{YORP}=1.0$.}
\label{fig.newagerho}
\end{center}
\end{figure}

While best-fit age has minimal dependence on $V_0$ and $c_{YORP}$, the
assumed value for density and thermal conductivity induce large
changes in the final age determination.  We show in Figure
\ref{fig.newagerho} the $\chi^2$ map of density vs. age for
simulations using $V_0=10~$m/s and $c_{YORP}=1.0$.  We note that for
$\rho=1.3~$g cm$^{-3}$ \citep[the value assumed by][]{bottke07} the
best fit age is $\sim80~$Myr which is consistent with the inverse
relation between age and the square root of the assumed albedo as
specified by those authors.  For a reasonable range of assumed
densities of $1.6-2.8~$g cm$^{-3}$ we find the best fitting age can
vary from $140-320~$Myr.  Without a better constraint on family member
density it will be difficult to more precisely determine the age of the
family.

\begin{figure}[ht]
\begin{center}
\includegraphics[scale=0.6]{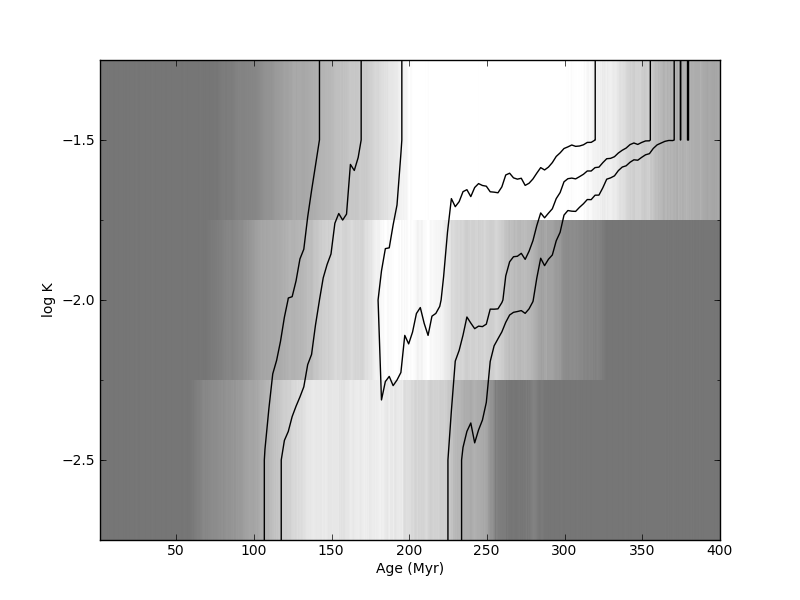} \protect\caption{The
  same as Figure \ref{fig.newagerho} but showing the log of thermal
  conductivity vs. age.}
\label{fig.newageK}
\end{center}
\end{figure}

Figure \ref{fig.newageK} shows a similar test, but now for a varied
thermal conductivity in the range of $0.003<K<0.03~$W m$^{-1}$
K$^{-1}$ (assuming $\rho=2.2~$g cm$^{-3}$).  Larger values of thermal
conductivity result in an increase in the best-fit age comparable to
the change caused by a larger assumed density.  Like density, the fact
that thermal conductivity is relatively unconstrained sets a
fundamental limit on the accuracy of simulations of family evolution
and age.

\begin{figure}[ht]
\begin{center}
\includegraphics[scale=0.6]{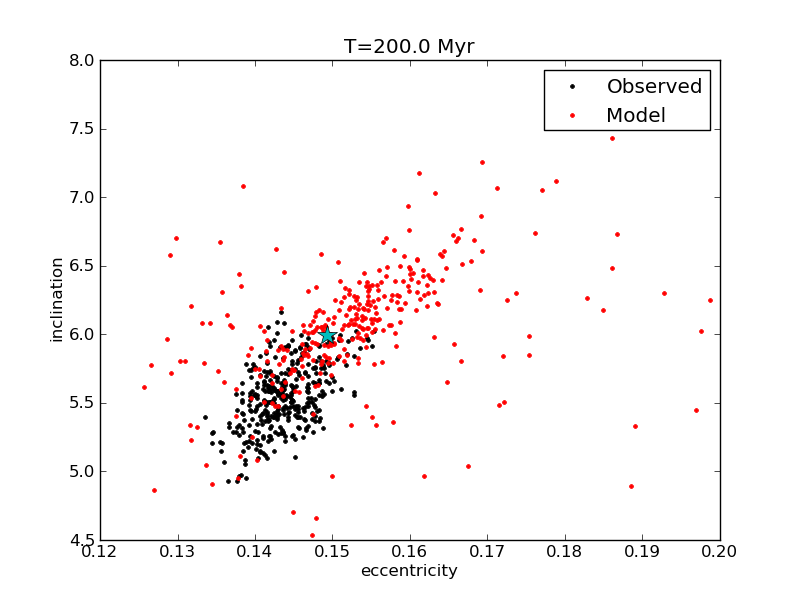} \protect\caption{Proper
  inclination (in degrees) vs. proper eccentricity for the observed
  Baptistina family members used for our analysis (black) and the
  simulated model family (red) using the best-fit values of
  $V_0=10~$m/s, $c_{YORP}=1$, and $T=200~$Myr and assumed values of
  $\rho=2.2~$g cm$^{-3}$ and $K=0.01~$W m$^{-1}$ K$^{-1}$.  The cyan
  star indicates the location of the parent body of the family.}
\label{fig.EIbest}
\end{center}
\end{figure}

\begin{figure}[ht]
\begin{center}
\includegraphics[scale=0.6]{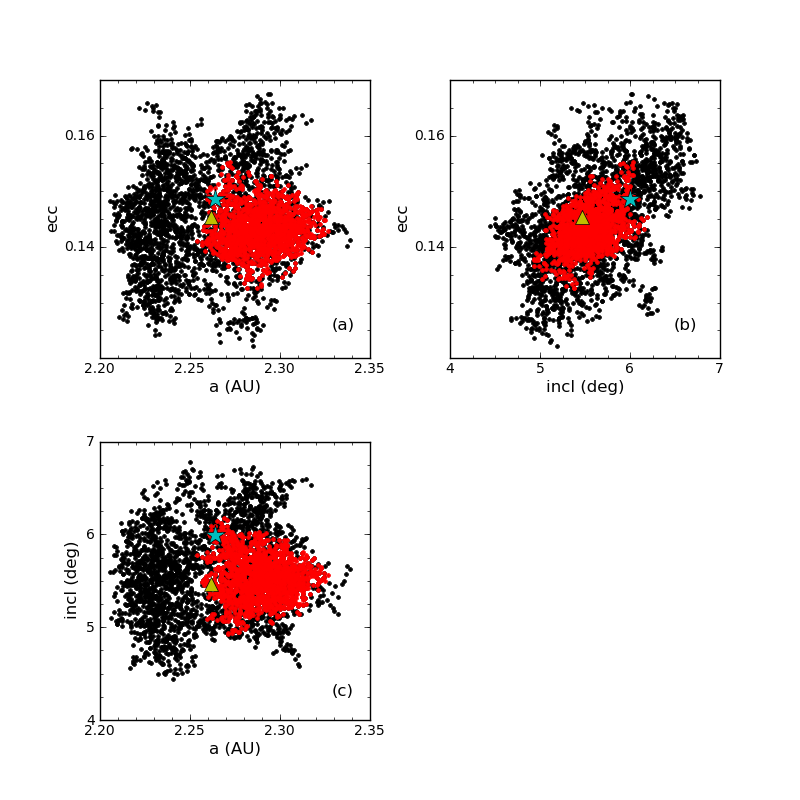} \protect\caption{Proper
  orbital elements for all members of the Baptistina collisional
  family in black, the restricted family list used for comparison to
  our simulations in red, (298) Baptistina as the cyan star, and
  (1696) Nurmela as the yellow triangle.}
\label{fig.BapAEI}
\end{center}
\end{figure}

We note that while our simulations can reproduce the semimajor axis
distribution of the family well for a variety of assumed parameters,
there are shortcomings to our solution.  In particular, we are unable
to simulate the observed distribution of the family members in
inclination-eccentricity space for any of the range of parameters
tested above.  We show in Figure~\ref{fig.EIbest} the
inclination-eccentricity distribution for the observed Baptistina
family compared with the family simulated using $V_0=10~$m/s,
$c_{YORP}=1$, $\rho=2.2~$g cm$^{-3}$, $K=0.01~$W m$^{-1}$ K$^{-1}$,
and an age of $T=200~$Myr.  Proper orbital elements are calculated for
the simulated particles using a frequency modified Fourier transform
\citep[FMFT][]{sidlichovsky96} with frequency filters described by
\citet{broz06}. The offset observed between the two populations may
indicate that the breakup had an ejection velocity distribution that
was highly anisotropic (unlike the assumed isotropic distribution used
in our simulations), that the assumed initial orbital parameters for
the parent at the time of breakup are incorrect, or that the asteroid
identified as the parent body is not the source of the breakup that
created the family.  As an example we show in Figure~\ref{fig.BapAEI}
the proper orbital elements of all objects identified as members of
the Baptistina family by \citet{nesvornyPDS}, the restricted list we
use for comparisons to our simulations (as discussed in
Section~\ref{sec.members}), (298) Baptistina, and (1696) Nurmela: the
largest body at the center of the a-e-i distribution which has a
diameter of $D=9.9~$km.  Future work will investigate these scenarios.

\section{Conclusions}
\label{sec.conc}

Using a symplectic integrator modified to include the effects of
gravity, Yarkovsky, and YORP, we have simulated the evolution of a
synthetic Baptistina asteroid family from breakup through
$\sim400~$Myr of evolution.  We compare the distribution at each
timestep to the observed distribution of the Baptistina family members
to determine the age of the family.  By varying all assumed
parameters, we set constraints on the effect of each parameter on the
determined age, and thus the error induced by the uncertainty in its
true value.

We find that while most physical parameters do not significantly
change our results, both the density and thermal conductivity of the
surface can drastically change the best-fit ages resulting in
uncertainties greater than $\sim50\%$, either younger or older.  While
having updated values for diameter and albedo reduces the uncertainty
in the simulation and the resultant age when compared to models
conducted using only absolute magnitude, assumptions for the other
physical parameters remain a significant source of uncertainty in the
calculation.

Using the WISE-derived albedos and diameters we find a best-fitting
age for the Baptistina family of $190\pm30~$Myr when we used a single
assumed density of $\rho=2.2~$g cm$^{-3}$ and an assumed thermal
conductivity of $K=0.01~$W m$^{-1}$ K$^{-1}$.  When we allow density
and thermal conductivity to vary over nominal ranges ($\pm30\%$, and
up or down by a factor of 3, respectively) and we find that the
best-fitting age can range anywhere from $140-320~$Myr.

The differences between our results and the findings of
\citet{bottke07} are due primarily to the smaller size of the
Baptistina family members that we measure compared to their assumed
values and the increase in the assumed density.  A higher assumed
density will weaken the non-gravitational forces compared to
gravitational perturbation and slow the overall evolution when strong
gravitational interactions do not dominate the process.  We also note
that the revised albedo and diameter measurements result in a
reduction in both the size of the pre-impact body and the number of
large fragments produced in the impact, decreasing the number
available to enter the near-Earth population.

Our simulations all assume that (298) Baptistina is the parent of the
Baptistina family and that its orbital elements at the time of breakup
were the same as today.  If instead a different object is the parent
of the family, then the family age may change dramatically from the
values found here.  A new suite of simulations would be required,
using the updated parent, to determine the family age.  Future work
will explore this possibility for the Baptistina family.

In the end, we are unable to set a firm constraint on the age of the
Baptistina family without more information about the family's physical
parameters ($\rho$ and $K$, specifically).  However, the uncertainty
in this age determination can be greatly reduced with focused
investigations of the family members.  In particular, thermophysical
modeling of a selection of Baptistina family members will allow us to
better constrain the physical parameters such as thermal conductivity,
while identification and study of any binary asteroids that may be
family members will allow us decrease the uncertainty in the density
of those bodies, and by extension the family as a whole.  Future work
will extend our investigation to the remaining asteroid families
observed by WISE, taking into account the caveats and concerns we
highlight here.

\section*{Acknowledgments}

We thank the referee, Bill Bottke, for his helpful and insightful
comments that resulted in a critical reanalysis of the data, greatly
improving our results and the manuscript in general.  We also thank
Bob McMillan for his editing of this manuscript.  J.R.M. was supported
by an appointment to the NASA Postdoctoral Program at JPL,
administered by Oak Ridge Associated Universities through a contract
with NASA.  Computer simulations for this research were carried out on
JPL's Zodiac supercomputer, which is administered by the JPL
Supercomputing and Visualization Facility.  The supercomputer used in
this investigation was provided by funding from the JPL Office of the
Chief Information Officer.  This publication makes use of data
products from the Wide-field Infrared Survey Explorer, which is a
joint project of the University of California, Los Angeles, and the
Jet Propulsion Laboratory/California Institute of Technology, funded
by the National Aeronautics and Space Administration.  This
publication also makes use of data products from NEOWISE, which is a
project of the Jet Propulsion Laboratory/California Institute of
Technology, funded by the Planetary Science Division of the National
Aeronautics and Space Administration.  This research has made use of
the NASA/IPAC Infrared Science Archive, which is operated by the Jet
Propulsion Laboratory, California Institute of Technology, under
contract with the National Aeronautics and Space Administration.


\end{document}